\documentclass[floatfix,aps,pra,groupedaddress,superscriptaddress,preprintnumbers,twocolumn]{revtex4}
\usepackage{graphicx}
\usepackage{bm}
\usepackage{amssymb,amsmath}

\begin{document}
	
	
	\title{Dynamical enhancement of nonparaxial effects \\in the electromagnetic field of a vortex electron}
	
	\author{Dmitry Karlovets} 
\affiliation{Faculty of Physics, Tomsk State University, Lenina Ave.\,36, 634050 Tomsk, Russia}
	

	\date{\today}
	
	
\begin{abstract}
A quantum state of an electron influences its electromagnetic field. If a spatial profile of the electron wave packet is not Gaussian,
the particle may acquire additional intrinsic multipole moments, which alter its field, especially at small distances.
Here the fields of a vortex electron with orbital angular momentum $\ell$ are obtained in a form of a multipole expansion 
with an electric quadrupole term kept by using the generalized (non-paraxial) Laguerre-Gaussian beams. 
The quadrupole contribution arises beyond a paraxial approximation, is linearly enhanced for highly twisted packets with $|\ell| \gg 1$, 
and can be important for the interactions of twisted beams with bulk matter and artificial structures. 
Moreover, this term results in an azimuthal asymmetry of the magnetic field in a rest frame of the electron,
which appears thanks to the spreading of the packet with time.
Thus, somewhat contrary to physical intuition, the spreading may enhance non-paraxial phenomena.
For the available electron beams, this asymmetry can in principle be reliably detected, which would be the first experimental evidence of a non-paraxial effect with the vortex electrons.
\end{abstract}


\maketitle

\section{Introduction}

Electrons with a definite projection of orbital angular momentum (OAM) onto a propagation axis -- the so-called vortex or twisted electrons -- were predicted theoretically \cite{Bliokh, Review} and recently obtained experimentally \cite{Uchida, Verbeeck, McMorran}. In the majority of cases, quantum states of such electrons can be described within a model of a so-called Bessel beam, 
which has a definite energy $\varepsilon$, a longitudinal momentum $p_z$, an absolute value of a transverse momentum $p_{\perp}$, a spin $s_z$, and the OAM $l_z \equiv \ell$.
Similar to a plane wave, this state is not localized in space and that is why in the problems for which the localization is crucial it needs to be replaced with a more elaborated model.

A current density of the Bessel beam does not depend on time and, therefore, an electromagnetic field of such an electron in a laboratory frame of reference is static.
However the field of a real moving electron is anything but static, whatever spatial profile the wave packet has.
The fields of the Bessel electron beam have been obtained in Ref.\cite{Lloyd}, but they do not actually coincide with those of a real vortex electron packet,
whose centroid is localized in space at a given moment of time. A physically consistent way to obtain the field of a vortex electron
is to take a spatially-localized wave packet, which represents an exact (or paraxial) solution to the Dirac equation, 
to calculate its current exactly, and then to employ a standard multipole expansion. The field of the electron will then represent a sum of those of the electron's multipole moments.
It is these OAM-induced intrinsic multipole moments that make the field of the vortex electron different from that of the ordinary OAM-less one
and it is this approach that I pursue in this paper. For the Bessel beam, the moments higher than the magnetic dipole one do not vanish but \textit{diverge} (see \cite{Multi} and below) 
because of the lack of localization, which makes it impossible to evaluate the field of a vortex electron within this model.

While the current and the electromagnetic field can be calculated in the laboratory frame, it is much easier to obtain them in the electron's rest frame first and then, 
given that the fields are transformed as components of a second rank tensor, to transform them into the former frame.
The paraxial Laguerre-Gaussian (LG) beams (see, for instance, Ref.\cite{Review}) cannot be used for such a task either, as they are restricted by the condition of paraxiality, $p_z \gg p_{\perp}$,
and are not applicable in the rest frame with $p_z = 0$. As I argue in Ref.\cite{PRA}, these beams can actually be used only for relativistic electrons but hardly for those with $\varepsilon_c \sim 300$ keV.

The generalized Laguerre-Gaussian beams, proposed in \cite{PRA}, can be used beyond the paraxial regime and, in particular, stay valid in the rest frame. 
They represent an exact solution to the Dirac equation in relativistic case and to the Schr\"odinger equation for non-relativistic energies. 
That is why I shall use this model to derive the intrinsic multipole moments of the vortex electron and to obtain its electromagnetic field.

An electric quadrupole moment and higher moments vanish for a rotationally symmetric (say, Gaussian) packet, but are finite for the vortex electron \cite{Multi}.
As I demonstrate hereafter, the former moment arises beyond the paraxial approximation only, violates the azimuthal symmetry and diverges with time as the packet spreads.
At some moment of time, $t \lesssim t_d$, the quadrupole contribution ceases to be small at all and its influence on the electron's field becomes easily noticeable.
To be precise, the magnetic field of the electron becomes azimuthally asymmetric even in the rest frame and this non-paraxial effect vanishes for the OAM-less beams or if the packet dynamics is neglected.

The non-paraxial phenomena can be enhanced for highly twisted electrons with $|\ell| \gg 1$ \cite{PRA}, and it turns out that, somewhat contrary to intuition, 
the spreading may further enhance some of them. The direct detection of the above azimuthal asymmetry is feasible with the already available electron beams, 
which would be the first observation of a non-paraxial effect with the vortex electrons. Along with the fundamental interest, these phenomena may affect the radiation and scattering processes 
with the twisted beams in matter and in electromagnetic fields at relatively low frequencies. A system of units $e = \hbar = c = 1$ is used.

\section{Generalized Laguerre-Gaussian beams}

The generalized Laguerre-Gaussian packets of a non-relativistic electron \cite{PRA},
\begin{widetext}
\begin{eqnarray}
& \displaystyle \psi_{\ell, n}({\bm r},t) = \sqrt{\frac{n!}{(n + |\ell|)!}} \frac{i^{2n+\ell}}{\pi^{3/4}}\frac{\rho^{|\ell|}}{(\sigma_{\perp}(t))^{|\ell| + 3/2}}\ L_{n}^{|\ell|}\left(\frac{\rho^2}{(\sigma_{\perp}(t))^2}\right) \exp\Big\{-it\langle p\rangle^2/2m + i\langle p\rangle z + i\ell\phi_r - \cr
& \displaystyle - i (2n + |\ell| + 3/2)\arctan (t/t_d) - \frac{1}{2(\sigma_{\perp}(t))^2}\, (1-it/t_d)\left(\rho^2 + (z-\langle u\rangle t)^2\right)\Big\},\cr
& \displaystyle \int d^3 r\, |\psi_{\ell, n}({\bm r},t)|^2 = 1,
\label{LGpsi}
\end{eqnarray}
\end{widetext}
represent an exact solution to the Schr\"odinger equation. 
Here,
\begin{eqnarray}
& \displaystyle (\sigma_{\perp}(t))^2 = \frac{1}{\sigma^2} \left(1 + t^2/t_d^2\right) = (\sigma_{\perp}(0))^2 + \left(\frac{\sigma}{m}\right)^2 t^2,\cr 
& \displaystyle \sigma_{\perp}(0) = \frac{1}{\sigma},\ t_d = \frac{m}{\sigma^2},\ \langle u\rangle = \frac{\langle p\rangle}{m} \ll 1.
\label{spread}
\end{eqnarray}
Clearly, spreading of the packet with time represents a non-paraxial effect, as it is attenuated by the following small parameter
\begin{eqnarray}
& \displaystyle \left(\frac{\sigma}{m}\right)^2 = \left(\frac{\lambda_c}{\sigma_{\perp}(0)}\right)^2 = |\ell|\left(\frac{\lambda_c}{\langle\rho(0)\rangle}\right)^2 \ll 1,
\label{param}
\end{eqnarray}
where
$$
\lambda_c \approx 3.9\times 10^{-11}\, \text{cm} 
$$
is the electron Compton wavelength and
\begin{eqnarray}
& \displaystyle
\langle \rho(t) \rangle = \sqrt{|\ell|}\,\sigma_{\perp}(t) = \frac{\sqrt{|\ell|}}{\sigma}\sqrt{1 + t^2/t_d^2}
\label{rho}
\end{eqnarray}
is a mean radius of the vortex packet. For available beams, the parameter (\ref{param}) does not exceed $10^{-6}$ \cite{Angstrom}, 
although the physical parameter that governs the non-paraxial corrections to observables is $|\ell|$ times larger than (\ref{param}) \cite{PRA}.

The diffraction time $t_d = m/\sigma^2$ in (\ref{spread}) can also be represented as follows:
\begin{eqnarray}
\displaystyle
t_d = t_c \left(\frac{\sigma_{\perp}(0)}{\lambda_c}\right)^2 \gg t_c,\ t_c = \lambda_c/c \approx 1.3\times 10^{-21}\, \text{sec.}
\label{td}
\end{eqnarray}
Although the LG packet (\ref{LGpsi}) is non-relativistic, it correctly describes the non-paraxial effects, which are closely connected to such a relativistic phenomenon as the appearance of antiparticles. 
Indeed, the time $t_c$ represents a characteristic lifetime of an electron-positron pair ($1/t_c = m$),
and the smallness of the parameter (\ref{param}) simply means that one cannot focus a one-electron wave packet to a spot smaller than the Compton wavelength without creation of the electron-positron pairs (see, for instance, Sec.1 in \cite{BLP}).

The state (\ref{LGpsi}) can also be obtained from a relativistic paraxial solution to the Klein-Gordon equation $\psi_{\ell,n}^{\text{par}}(x)$ as follows (see Eq.(51) in \cite{PRA}):
\begin{eqnarray}
& \displaystyle \psi_{\ell,n}({\bm r},t) = \sqrt{2m}\,\,\psi_{\ell,n}^{\text{par}}(x)\Big|_{\langle p\rangle \ll m}e^{i m t}.
\label{LGfromrel}
\end{eqnarray} 
It comes of no surprise that a non-paraxial but non-relativistic exact solution can be obtained from a paraxial -- that is, approximate -- relativistic one (see, for instance, \cite{Bagrov}). 

Note that upon time inversion we have
\begin{eqnarray}
& \displaystyle t \rightarrow -t:\quad \ell \rightarrow - \ell,\ \langle p \rangle \rightarrow - \langle p\rangle,\ z \rightarrow z,\ \phi_r \rightarrow \phi_r,
\label{tinv}
\end{eqnarray} 
and so
\begin{eqnarray}
& \displaystyle t \rightarrow -t:\quad \psi_{\ell, n}({\bm r},t) \rightarrow \psi_{\ell, n}^*({\bm r},t),
\label{tinv2}
\end{eqnarray}
as should be according to the general principles of quantum mechanics and, in particular, of relativistic CPT-invariance \cite{BLP}. 
I emphasize that it is a $\mathcal T$-odd \textit{time-dependent Gouy phase},
\begin{eqnarray}
& \displaystyle \arctan (t/t_d),
\label{Gouy}
\end{eqnarray} 
that provides the correct transformation (\ref{tinv2}) of the wave function under the time inversion. 
The customary paraxial LG beams with the $\mathcal T$-even Gouy phase depending on the distance $z$,
\begin{eqnarray}
& \displaystyle \arctan (z/z_R),
\label{Gouyz}
\end{eqnarray} 
do not provide such a correct transformation and, therefore, \textit{violate the CPT-invariance}, which is highly problematic for a consistent relativistic theory of an electron.

\section{Intrinsic multipole moments}

For a scalar non-relativistic packet with a mass $m$ and the wave function $\psi ({\bm r},t)$, the components of the current $j^{\mu} = \{j^0, {\bm j}\}$ are
\begin{eqnarray}
& \displaystyle j^0 ({\bm r},t) = |\psi ({\bm r},t)|^2,\ \int d^3 r\, j^0 ({\bm r},t) = 1,\cr 
& \displaystyle {\bm j} ({\bm r},t) = \psi^*({\bm r},t)\frac{-i}{2m}{\bm\nabla}\psi({\bm r},t) + \text{c.c.}
\label{j}
\end{eqnarray}
Given the definitions of the first three multipole moments,
\begin{eqnarray}
& \displaystyle {\bm d}(t) = \int d^3 r\, {\bm r}\, j^0 ({\bm r},t),\ \ {\bm \mu}(t) = \frac{1}{2}\int d^3r\, {\bm r}\times {\bm j} ({\bm r},t),\cr
& \displaystyle Q_{\alpha\beta}(t) = \int d^3r\, j^0 ({\bm r},t) \left (3 r_{\alpha}r_{\beta} - {\bm r}^2 \delta_{\alpha\beta}\right),
\label{dmu}
\end{eqnarray}
the corresponding intrinsic values are \cite{Multi}
\begin{eqnarray}
& \displaystyle {\bm d}_{\text{int}} = 0,\ {\bm \mu}_{\text{int}}(t) = {\bm \mu}(t) - \frac{1}{2}\int d^3r\, {\bm d}(t)\times {\bm j} ({\bm r},t),\cr
& \displaystyle Q_{\alpha\beta,\text{int}}(t) = Q_{\alpha\beta}(t)-3d_{\alpha}(t)d_{\beta}(t) + {\bm d}^2(t) \delta_{\alpha\beta},\cr 
& \displaystyle \alpha,\beta = 1,2,3.
\label{dmuint}
\end{eqnarray}
In what follows, I deal with the intrinsic moments only and omit the subscript ``int''.

Let me start in a frame of reference in which the packet is at rest on average, 
$$
\langle {\bm u}\rangle = 0.
$$ 
For the fundamental mode with $n=0$ of the LG packet (\ref{LGpsi}), the charge density and the current density are
\begin{eqnarray}
& \displaystyle j_{\ell}^{0}({\bm r},t) = \frac{1}{\pi^{3/2}|\ell|!}\, \frac{\rho^{2|\ell|}}{(\sigma_{\perp}(t))^{2|\ell|+3}}\, \exp\left\{-\frac{{\bm r}^2}{(\sigma_{\perp}(t))^2}\right\},\cr
& \displaystyle {\bm j}_{\ell}({\bm r},t) = j_{\ell}^{0}({\bm r},t) \left(\frac{{\bm r}t}{t^2 + t_d^2} + {\bm e}_{\phi} \frac{\ell}{m\rho}\right),\cr
& \displaystyle \text{and}\quad \partial_{\mu} j_{\ell}^{\mu} = 0,
\label{LGcurrent}
\end{eqnarray} 
where
\begin{eqnarray}
& \displaystyle {\bm e}_{\phi} = \{-\sin \phi, \cos \phi, 0\},\cr 
& \displaystyle {\bm e}_{\rho} = {\bm \rho}/\rho =\{\cos\phi,\sin\phi,0\},\ \hat{\bm z} = \{0,0,1\},\cr 
& \displaystyle {\bm r} = r {\bm n} = r \{\sin\theta \cos \phi, \sin\theta\sin \phi, \cos\theta\} = \cr
& \displaystyle = \{{\bm \rho}, z\} = \{\rho \cos \phi, \rho \sin \phi, z\} = \rho {\bm e}_{\rho} + z \hat{\bm z}.
\label{not}
\end{eqnarray} 
Note that the charge density does not depend on the azimuthal angle $\phi$. Recall that for a classical particle the current $j^{\mu}_{\ell} =\{j^0_{\ell}, {\bm j}_{\ell}\}$ represents a time-like four-vector,
$$
{\bm j}_{\ell}({\bm r},t) = j_{\ell}^{0}({\bm r},t) \langle{\bm u}\rangle.
$$
This is not the case for the ``quantum'' current density ${\bm j}_{\ell}({\bm r},t)$ from Eq.(\ref{LGcurrent}), which does not vanish in the rest frame. 
Moreover, this current has all three components,
\begin{eqnarray}
&& \displaystyle {\bm j}_{\ell}({\bm r},t) = {\bm e}_{\phi} j_{\phi} + {\bm e}_{\rho} j_{\rho} + \hat{\bm z} j_z,
\label{LGcurrent2}
\end{eqnarray} 
whereas the corresponding current of the Bessel beam,
\begin{eqnarray}
&& \displaystyle {\bm j}_{\ell}({\bm r},t) \equiv {\bm j}_{\ell}(\rho, \phi) = j_{\ell}^{0}({\rho})\, {\bm e}_{\phi} \frac{\ell}{m\rho},\cr
&& \displaystyle j_{\ell}^{0}({\rho}) = N^2 J_{\ell}^2 (p_{\perp}\rho),\ N = \text{const},
\label{Besscurrent}
\end{eqnarray} 
has only an azimuthal component in the rest frame. Clearly, when the LG packet is wide, $\sigma_{\perp}(0) \rightarrow \infty, t \gg t_d$, 
the current (\ref{LGcurrent}) becomes similar to that of the Bessel beam\footnote{This is in accord with the fact that the very Bessel beam represents merely a special case of the Laguerre-Gaussian packet, 
obtained from the latter when $\sigma_{\perp}(0) \rightarrow \infty, n \rightarrow \infty$ \cite{PRA}.}. 

However for any finite moment of time, $t \lesssim t_d$, the term 
\begin{eqnarray}
&& \displaystyle
{\bm j}_{\ell}({\bm r},t) \propto j_{\ell}^{0}({\bm r},t)\,\frac{{\bm r}t}{t^2 + t_d^2}
\label{Current1}
\end{eqnarray}
violates an explicit $\mathcal T$-invariance of the current, which happens because of the packet's spreading with time and not due to the OAM,
as this feature holds even when $\ell = 0$. 

Substituting the current (\ref{LGcurrent}) into Eq.(\ref{dmuint}), we arrive at the following intrinsic moments \footnote{In order to take into account the spin, 
one would make the following substitution in ${\bm \mu}$: $\ell \rightarrow \ell + 2s_z,\, s_z = \pm 1/2$. 
However, the non-paraxial corrections to the magnetic moment are of the order of $|\ell|\,\lambda_c^2/(\sigma_{\perp}(0))^2$ \cite{PRA} 
and, therefore, they can compete with the quadrupole contribution. I neglect this spin-orbit interaction effect by considering only \textit{unpolarized} beams.}:
\begin{eqnarray}
& \displaystyle {\bm \mu} = \frac{\ell}{2m}{\hat{\bm z}},\ Q_{\alpha\beta}(t) = \langle \rho(t) \rangle^2\, \text{diag}\{1/2,1/2,-1\}.
\label{Moments}
\end{eqnarray} 
Clearly, as the packet spreads the quadrupole moment grows with time. However, for the OAM-less Gaussian beam this moment vanishes because $\langle \rho(t) \rangle^2 \propto |\ell|$.

The coordinate-momentum uncertainties for this state are \cite{PRA}
\begin{eqnarray}
& \displaystyle \Delta x \Delta p_x = \Delta y \Delta p_y = \frac{1}{2} (|\ell| + 1) \sqrt{1 + t^2/t_d^2}
\label{unc}
\end{eqnarray} 
and a number of the quantum states in a transverse phase space
\begin{eqnarray}
& \displaystyle \Delta \Gamma = \frac{\Delta x \Delta p_x \Delta y \Delta p_y}{(2\pi)^2} \propto (|\ell| + 1)^2 (1 + t^2/t_d^2)
\label{density}
\end{eqnarray} 
grows with time together with a corresponding entropy $S$,
\begin{eqnarray}
& \displaystyle S = \ln \Delta \Gamma  \propto \ln (|\ell| + 1)^2 (1 + t^2/t_d^2).
\label{entropy}
\end{eqnarray} 

Next, it is because of the packet's spreading that the corresponding solution of the Maxwell equations has a sense only for not very large times,
$$
t \lesssim t_d.
$$
Indeed, at $|t| \gg t_d$ the packet becomes unlocalized in space, $\langle \rho(t) \rangle \rightarrow \infty$ (akin to the Bessel beam), 
and the very multipole expansion, applicable when $r \gtrsim \langle \rho(t) \rangle$, loses its sense.

Importantly, the third derivative over time of the quadrupole moment vanishes, 
$$
\dddot{Q}_{\alpha\beta}(t) = 0,
$$ 
and so this time dependence, which is closely connected to the packet's spreading and to the increase of the entropy, does not lead to the radiation of electromagnetic waves 
because the radiation intensity is proportional to $\dddot{Q}_{\alpha\beta}(t)$ (see Sec.\,71 in Ref.\cite{LL2}). 
Given that the magnetic moment does not depend on time, the radiation fields, which decay as $|{\bm E}^R| \propto 1/r, |{\bm H}^R| \propto 1/r$, 
simply vanish in all orders of the multipole expansion, as expected for a freely propagating particle. 
Therefore, we need to evaluate only the non-radiating (evanescent) fields, $|{\bm E}|,|{\bm H}| \propto 1/r^k,\, k=2,3,4$.
For this purpose, one can use the corresponding multipole expansion of the retarded potentials (see the problem 1 of Sec.\,72 in Ref.\cite{LL2}) \footnote{When deriving the multipole expansion in \cite{LL2}, the current was taken to be classical, ${\bm j} = j^0 {\bm u}$, and which is not the case for the vortex packet, as Eq.(\ref{LGcurrent}) demonstrates. It can be easily shown, however, that the very same formulas connecting the retarded potentials $A^0, {\bm A}$ and the multipole moments ${\bm \mu}, Q_{\alpha\beta}$ stay valid for the currents with ${\bm j} \ne j^0 {\bm u}$ as well. In particular, it is the case for the LG beam whose azimuthal component of the current $j_{\phi}$ contributes only to ${\bm \mu}$, but not to $Q_{\alpha\beta}$.}.

\section{Fields in the rest frame}

The fields of the vortex electron represent a sum of those of the charge $e$, of the magnetic moment ${\bm \mu}$, and of the electric quadrupole moment $Q_{\alpha\beta}$.
In the rest frame, they are
\begin{eqnarray}
& \displaystyle {\bm E}({\bm r},t) = {\bm E}_e({\bm r}) + {\bm E}_Q({\bm r},t),\cr
& \displaystyle {\bm H}({\bm r},t) = {\bm H}_{\mu}({\bm r}) + {\bm H}_Q({\bm r},t),\cr 
& \displaystyle {\bm E}_e({\bm r}) = \frac{{\bm n}}{r^2},\ {\bm H}_{\mu}({\bm r}) = \frac{3{\bm n}({\bm n}\cdot{\bm \mu}) - {\bm \mu}}{r^3},\cr
& \displaystyle {\bm E}_Q ({\bm r},t) = \frac{5}{2}\,{\bm n} \frac{({\bm n}\cdot{\bm Q})}{r^4} - \frac{{\bm Q}}{r^4} + \frac{5}{2}\,{\bm n} \frac{({\bm n}\cdot{\dot{\bm Q}})}{r^3} - \cr
& \displaystyle - \frac{\dot{\bm Q}}{r^3} + {\bm n} \frac{({\bm n}\cdot\ddot{\bm Q})}{r^2} - \frac{\ddot{\bm Q}}{2 r^2},\cr
& \displaystyle {\bm H}_Q ({\bm r},t) = - \frac{1}{2r^3}\, {\bm n}\times \dot{\bm Q} -\frac{1}{2r^2}\, {\bm n}\times\ddot{\bm Q},
\label{EHfar}
\end{eqnarray} 
where 
\begin{eqnarray}
& \displaystyle {\bm Q}_{\alpha} \equiv {\bm Q}_{\alpha}(t-r) = Q_{\alpha\beta}(t-r)n_{\beta} = \cr
& \displaystyle = \langle \rho(t-r) \rangle^2 \left\{\frac{1}{2}\sin\theta\cos\phi,\frac{1}{2}\sin\theta\sin\phi,-\cos\theta\right\}_{\alpha},
\label{qn}
\end{eqnarray} 
the dots mean derivatives over time, and all the values in the right-hand side are taken at the retarded moment of time, $t-r$.
I emphasize that this expression is applicable not only in the wave zone with $r \gg \langle \rho (t) \rangle$, 
but also not too far from the source, 
$$
r \gtrsim \langle \rho (t) \rangle,
$$
where only the static fields exist and the terms with $\dot{\bm Q}, \ddot{\bm Q}$ in ${\bm E}_Q$ can be neglected.
Finally, the fields inside the vortex core, at $r < \langle \rho (t) \rangle$, can be found by the numerical integration of the retarded potentials.

After some algebra, the fields in the cylindrical coordinates become
\begin{widetext}
\begin{eqnarray}
&& \displaystyle {\bm E}_e({\bm r}) =  E_{e,\rho}\, {\bm e}_{\rho} + E_{e,z}\, \hat{\bm z},\ E_{e,\rho} = \frac{\sin \theta}{r^2},\ E_{e,z} = \frac{\cos \theta}{r^2},\ E_{e,\phi} = 0,\cr
&& \displaystyle {\bm H}_{\mu}({\bm r}) =  H_{\mu,\rho}\, {\bm e}_{\rho} + H_{\mu,z}\, \hat{\bm z},\ H_{\mu,\rho} = \frac{\ell}{2m}\,\frac{3\sin\theta\cos\theta}{r^3},\ H_{\mu,z} = \frac{\ell}{2m}\frac{3\cos^2\theta-1}{r^3},\ H_{\mu,\phi} = 0,\cr
&& \displaystyle {\bm E}_Q ({\bm r},t) = E_{Q,\rho}\, {\bm e}_{\rho} + E_{Q,z}\, \hat{\bm z},\ E_{Q,\phi} = 0,\cr
&& \displaystyle E_{Q,\rho}({\bm r},t) = \frac{\sin\theta}{4r^2} \left(3\frac{\langle\rho(0)\rangle^2}{r^2} (1-5\cos^2\theta) + \ell^2\left(\frac{\lambda_c}{\langle\rho(0)\rangle}\right)^2 \left[3\left(\frac{t}{r}\right)^2 (1-5\cos^2\theta) + 3\cos^2\theta - 1\right]\right),\cr
&& \displaystyle E_{Q,z}({\bm r},t) = \frac{\cos\theta}{4r^2} \left(3\frac{\langle\rho(0)\rangle^2}{r^2} (3-5\cos^2\theta) + \ell^2\left(\frac{\lambda_c}{\langle\rho(0)\rangle}\right)^2 \left[3\left(\frac{t}{r}\right)^2 (3-5\cos^2\theta) + 3\cos^2\theta - 1\right]\right),\cr
&& \displaystyle {\bm H}_Q ({\bm r},t) = H_{Q,\phi}\,{\bm e}_{\phi} = -\frac{3}{2}\frac{t}{r}\frac{\ell^2}{r^2}\left(\frac{\lambda_c}{\langle\rho(0)\rangle}\right)^2\sin\theta\cos\theta\, {\bm e}_{\phi},\ H_{Q,\rho} = H_{Q,z} = 0,
\label{Fcomp}
\end{eqnarray}
\end{widetext} 
The quadrupole fields ${\bm E}_{Q}({\bm r},t)$ and ${\bm H}_{Q}({\bm r},t)$ \textit{grow linearly} with the OAM $|\ell|$, do not depend on its sign, 
and along with the ratio $\langle\rho(0)\rangle^2/r^2$, typical for the quadrupole contribution already at the classical level, 
they also contain the non-paraxial \textit{purely quantum} terms of the order of
$$
\ell^2 \left(\frac{\lambda_c}{\langle\rho(0)\rangle}\right)^2 = \mathcal O(\hbar^2),
$$
which are $|\ell|$ times enhanced compared to the small parameter of the problem, Eq.(\ref{param}). 

As a result, 
\begin{itemize}
\item
The non-paraxial terms decay slower with the distance than the quasi-classical ones;
\item
They are just moderately attenuated for highly twisted beams with $|\ell| \gg 1$, in accord with the more general analysis of non-paraxial corrections \cite{PRA};
\item
The azimuthal component of the magnetic field $H_{Q,\phi}$ is of the order of $\ell^2 t\, \lambda_c^2/\langle \rho(0)\rangle^2 = \mathcal O(|\ell|t)$ and is dynamically enhanced at large times $t\lesssim t_d$. 
\end{itemize}
For electron beams with the typical width of 
$$
\langle\rho(0)\rangle \sim 10\, \text{nm} - 100\, \mu \text{m} \gg \lambda_c,
$$ 
we stay well within the paraxial approximation, but the terms $\ell^2\lambda_c^2/\langle\rho(0)\rangle^2$ can be safely neglected only at small times, $t < r$, because of the packet dynamics.
In the paraxial regime, the magnetic field is azimuthally symmetric and both fields do not explicitly depend upon the sign of time.

Generically, the electric field ${\bm E}$ is even with respect to the time inversion, while the magnetic field ${\bm H}$ is odd. 
For the field ${\bm H}_{\mu}$ this is the case because the OAM $\ell$ changes its sign when $t \rightarrow - t$.
Beyond the paraxial regime, however, the field ${\bm H}_{Q}$ explicitly depends on the sign of time, which is closely
connected with the time dependence of the Gouy phase (9). This magnetic field acquires an azimuthal $H_{Q,\phi}$ component due to the spreading with time, and the effect vanishes for the ordinary OAM-less Gaussian beams or if we neglect the dynamics.

The magnitude of this non-paraxial effect can be quantified by the following ratio (\textit{an azimuthal asymmetry}):
\begin{eqnarray}
& \displaystyle
\mathcal A(t) = H_{\phi} ({\bm r},t)/H_{\rho}({\bm r},t) = H_{Q,\phi}/H_{\mu,\rho} = \cr
& \displaystyle = - \text{sign}(\ell) \left(\frac{\lambda_c}{\sigma_{\perp}(0)}\right)^2 \frac{t}{t_c},   
\label{Asymm}
\end{eqnarray}
which does not grow with the OAM but depends on its sign, is even under time inversion, and where the field components are measured at the same distance from the electron. 
Note that they both decay as $1/r^3$, which is slower than the classical quadrupole contribution, $1/r^4$. 
This asymmetry is proportional to the small parameter (\ref{param}), but is enhanced at large times. 

For highly twisted beams with $|\ell| \gg 1$, both the fields ${\bm H}_{\mu}$ and ${\bm H}_{Q}$ grow linearly with the OAM,
which makes their measurements easier. Indeed, the magnetic field of the spin magnetic moment is usually too weak to be noticeable, 
whereas the field of the OAM-induced magnetic moment is roughly $\ell$ times stronger and $\ell$ can already reach the values of $\ell \sim 10^3$ \cite{l1000}.  
When $\ell = 0$, however, the fields vanish and the asymmetry no longer has a sense \footnote{If we take the spin into account, we obtain $|\ell|/(\ell + 2s_z)$ instead of $\text{sign}(\ell)$, and the asymmetry simply vanishes for $\ell = 0$.}. 

Next, as the proper time $t$ is measured in the rest frame, the ratio (\ref{Asymm}) is Lorentz invariant for longitudinal boosts. 
The contribution of the higher multipole moments can be neglected only for $t \lesssim t_d = m/\sigma^2$,
and so the maximum value of the asymmetry in our approximation is
\begin{eqnarray}
& \displaystyle \mathcal |A| \lesssim \frac{t_d}{t_c}\, |\ell| \left(\frac{\lambda_c}{\langle\rho(0)\rangle}\right)^2 = 1,
\label{Asymmmax}
\end{eqnarray}
that is, the quadrupole contribution can become comparable with that of the magnetic moment,
\begin{eqnarray}
& \displaystyle |H_{Q,\phi}| \lesssim |H_{\mu,\rho}|.
\label{Asymmmax2}
\end{eqnarray}

Let's suppose that the electron packet is rather wide, $\langle\rho(0)\rangle \sim 10\, \mu\text{m}$, and that $|\ell| \sim 1$, which is typical for a beam of an electron microscope. 
Then we have
\begin{eqnarray}
&& \displaystyle
|\mathcal A| \sim \frac{t}{t_c\, 10^{16}} \approx \frac{t}{10^{-5}\, \text{sec.}}.   
\label{Aestim}
\end{eqnarray}
As a result, at $t \lesssim t_d \sim 10^{-5}\, \text{sec.}$ the azimuthal component of the magnetic field becomes easily noticeable. 
On the other hand, a $300$-keV beam would cover a distance of the order of $1$ km during this time, 
and so the measurements with the more tightly focused beams, $\langle\rho(0)\rangle \sim 1\, \text{nm} - 1\, \mu\text{m}$, seem to be preferable.

\section{Fields in the laboratory frame}

Let me now make a Lorentz boost to the laboratory frame in which the particle moves along the $z$ axis\footnote{In this paper, I treat only the longitudinal boosts which coincide with the direction of the OAM ($z$). The effects of the transverse boosts were studied in Ref.\cite{Bliokh17trans} for the Bessel beam.} with a velocity $\langle u \rangle \equiv \beta$ according to the law
$$
\langle z\rangle = \langle u \rangle t
$$
and with a Lorentz factor $\gamma = \langle\varepsilon\rangle/m = 1/\sqrt{1- \beta^2}$.
The fields in this frame are
\begin{eqnarray}
&& \displaystyle E^{(\text{lab})}_{\rho} = \gamma (E_{\rho} + \beta H_{\phi}) = \gamma (E_{e,\rho} + E_{Q,\rho} + \beta H_{Q,\phi}),\cr
&& \displaystyle E^{(\text{lab})}_{\phi} = \gamma (E_{\phi} - \beta H_{\rho}) = -\gamma \beta H_{\mu,\rho},\ E^{(\text{lab})}_z = E_z,\cr
&& \displaystyle H^{(\text{lab})}_{\rho} = \gamma (H_{\rho} - \beta E_{\phi}) = \gamma H_{\mu,\rho},\cr
&& \displaystyle H^{(\text{lab})}_{\phi} = \gamma (H_{\phi} + \beta E_{\rho}) = \gamma (H_{Q,\phi} + \beta E_{e,\rho} + \beta E_{Q,\rho}),\cr
&& \displaystyle H^{(\text{lab})}_z = H_z.
\label{FLabtrans}
\end{eqnarray} 
Simultaneously, we need to transform the coordinates, the angle $\theta$, and the time as follows:
\begin{eqnarray}
&& \displaystyle \rho = \text{inv},\ z\rightarrow \gamma (z - \beta t),\ t \rightarrow \gamma (t - \beta z),\cr
&& \displaystyle r^2 \rightarrow \rho^2 + \gamma^2 (z - \beta t)^2,\ \phi = \text{inv},\cr
&& \displaystyle \sin\theta \rightarrow \frac{\sin\theta}{\gamma (1-\beta\cos\theta)},\ \cos\theta \rightarrow \frac{\cos\theta - \beta}{1-\beta\cos\theta}.
\label{trans}
\end{eqnarray} 

\textit{The total} electric field, which includes the contributions of the charge, of the magnetic moment, and of the quadrupole moment, is
\begin{widetext}
\begin{eqnarray}
&& \displaystyle E^{(\text{lab})}_{\rho} = \frac{s}{1 - \beta c} \frac{1}{\rho^2 + \gamma^2 (z-\beta t)^2} \left(1 + \frac{1}{4}\frac{\langle\rho(0)\rangle^2}{\rho^2 + \gamma^2 (z-\beta t)^2}\, A_{\rho} + \frac{1}{4}\,\ell^2\left(\frac{\lambda_c}{\langle\rho(0)\rangle}\right)^2 B_{\rho}(t)\right),\cr
&& \displaystyle A_{\rho} = 3 \left(1 - 5 \frac{(c-\beta)^2}{(1-\beta c)^2}\right),\quad B_{\rho}(t) = 3\gamma^2 \frac{(t-\beta z)^2}{\rho^2 + \gamma^2 (z-\beta t)^2}\left(1 - 5 \frac{(c-\beta)^2}{(1-\beta c)^2}\right) -\cr
&& \displaystyle \qquad \qquad \qquad \qquad - 6\beta\gamma\frac{t - \beta z}{\sqrt{\rho^2 + \gamma^2 (z-\beta t)^2}}\frac{c - \beta}{1-\beta c} +3 \frac{(c-\beta)^2}{(1-\beta c)^2} - 1, \cr
&& \displaystyle E^{(\text{lab})}_{\phi} = -3\beta \frac{\ell}{2m} \frac{s (c-\beta)}{(1-\beta c)^2}\frac{1}{(\rho^2 + \gamma^2 (z-\beta t)^2)^{3/2}},\cr 
&& \displaystyle E^{(\text{lab})}_z = \frac{c-\beta}{1-\beta c} \frac{1}{\rho^2 + \gamma^2 (z-\beta t)^2} \left(1 + \frac{1}{4}\frac{\langle\rho(0)\rangle^2}{\rho^2 + \gamma^2 (z-\beta t)^2}\, A_{z} + \frac{1}{4}\,\ell^2\left(\frac{\lambda_c}{\langle\rho(0)\rangle}\right)^2 B_{z}(t)\right),\cr
&& \displaystyle A_z = 3 \left(3 - 5 \frac{(c-\beta)^2}{(1-\beta c)^2}\right), B_z(t) = 3\gamma^2 \frac{(t-\beta z)^2}{\rho^2 + \gamma^2 (z-\beta t)^2}\left(3 - 5 \frac{(c-\beta)^2}{(1-\beta c)^2}\right) + 3 \frac{(c-\beta)^2}{(1-\beta c)^2} - 1,
\label{ELab}
\end{eqnarray}
\end{widetext}
whereas the total magnetic field becomes
\begin{widetext}
\begin{eqnarray}
&& \displaystyle H^{(\text{lab})}_{\rho} = -\frac{1}{\beta}\, E^{(\text{lab})}_{\phi},\cr
&& \displaystyle H^{(\text{lab})}_{\phi} = \frac{s}{1 - \beta c} \frac{1}{\rho^2 + \gamma^2 (z-\beta t)^2} \left(\beta + \frac{1}{4}\frac{\langle\rho(0)\rangle^2}{\rho^2 + \gamma^2 (z-\beta t)^2}\, C_{\phi} + \frac{1}{4}\,\ell^2\left(\frac{\lambda_c}{\langle\rho(0)\rangle}\right)^2 D_{\phi}(t)\right),\cr 
&& \displaystyle C_{\phi} = \beta A_{\rho},\quad D_{\phi}(t) = 3\beta\gamma^2 \frac{(t-\beta z)^2}{\rho^2 + \gamma^2 (z-\beta t)^2}\left(1 - 5 \frac{(c-\beta)^2}{(1-\beta c)^2}\right) -\cr
&& \displaystyle \qquad \qquad \qquad \qquad - 6\gamma\frac{t - \beta z}{\sqrt{\rho^2 + \gamma^2 (z-\beta t)^2}}\frac{c - \beta}{1-\beta c} +3\beta \frac{(c-\beta)^2}{(1-\beta c)^2} - \beta,\cr
&& \displaystyle H^{(\text{lab})}_z = \frac{\ell}{2m} \left (3\frac{(c-\beta)^2}{(1-\beta c)^2} - 1\right )\frac{1}{(\rho^2 + \gamma^2 (z-\beta t)^2)^{3/2}},
\label{HLab}
\end{eqnarray}
\end{widetext} 
where I have denoted
$$
s \equiv \sin\theta,\ c \equiv \cos\theta.
$$

In order to obtain the asymmetry (\ref{Asymm}) in the laboratory frame one has also measure the components of the electric field,
\begin{eqnarray}
&& \displaystyle
\mathcal A (t,z) = \frac{H_{\phi}^{(\text{lab})} - \beta E_{\rho}^{(\text{lab})}}{H_{\rho}^{(\text{lab})} + \beta E_{\phi}^{(\text{lab})}} = \cr
&& \displaystyle = - \ell \left(\frac{\lambda_c}{\langle\rho(0)\rangle}\right)^2 \gamma\, \frac{t - \beta z}{t_c},   
\label{AsymmLab}
\end{eqnarray}
Recall that this ratio itself is a function of the total fields, as in Eqs.(\ref{ELab}),(\ref{HLab}), that it is Lorentz invariant for longitudinal boosts, and that $\gamma (t - \beta z)$ is just the proper time in the electron's rest frame. It might seem that by performing the measurements in the plane $z=0$ we get an enhancement of the asymmetry due to the factor $\gamma$. 
However, as the wave packet now moves according to the law $\langle z\rangle = \beta t$, it is experimentally convenient to measure the fields 
at the plane where the electron currently is, i.e. at $z = \langle z\rangle$. This brings about the following expression 
\begin{eqnarray}
&& \displaystyle
\mathcal A (t,\beta t) = - \ell \left(\frac{\lambda_c}{\langle\rho(0)\rangle}\right)^2 \frac{1}{\gamma}\, \frac{t}{t_c}, 
\label{AsymmLabz}
\end{eqnarray}
which simply illustrates the slowing down of time in a moving frame. In order to detect this asymmetry one should perform at least two sets of measurements:
$$
\text{at}\ t = 0, \langle z\rangle = 0\quad \text{and at}\quad t \lesssim t_d, \langle z\rangle \lesssim \beta t_d.
$$

During the diffraction time $t_d$, an electron with the energy of $\varepsilon_c \sim 300\, \text{keV}$ and $\beta \approx 0.78$ will cover the distance of
$$
z_d = \beta\, t_d = \beta \lambda_c \left (\frac{\sigma_{\perp}(0)}{\lambda_c}\right )^2,
$$
which yields 
\begin{eqnarray}
& \displaystyle
z_d \sim 1\, \text{mm}\quad \text{for}\ \langle\rho(0)\rangle \sim 10\, \text{nm},\cr 
& \displaystyle \text{or}\ z_d \sim 10\, \text{cm}\quad \text{for}\ \langle\rho(0)\rangle \sim 100\, \text{nm},\ |\ell| \sim 1.
\label{zestim}
\end{eqnarray}
That is why the optimal beam width for the registration of the asymmetry lies within the following interval: 
\begin{eqnarray}
&& \displaystyle
\langle\rho(0)\rangle \sim 1\, \text{nm} - 1\, \mu\text{m},
\label{widthopt}
\end{eqnarray}
which is easily achievable with an electron microscope. Simultaneously, one should \textit{not} necessarily strive to make the OAM as large as possible, as the asymmetry does not depend on its value.
On the other hand, the fields themselves do depend on it and they are more easily detectable for higher values of the OAM.
The electron beams with the moderately large OAM of $|\ell| \sim 10-100$ would most likely suffice.

\section{Discussion}

The electromagnetic field of an electron depends on a quantum state of the latter because, for non-Gaussian wave packets, the electron may acquire additional intrinsic multipole moments.
As I have shown, this is the case for the vortex electrons, which are also endowed with the electric quadrupole moment beyond the paraxial approximation.
As the packet spreads, this moment grows with time and results in an azimuthal asymmetry of the electron's magnetic field at large times (low frequencies).
Such an asymmetry arises from an interplay between the two purely quantum phenomena -- the packet's spreading and the possession of the OAM. 
If detected, this asymmetry would be the first non-paraxial effect measured with the vortex electron beams.

Along with the direct detection of the field asymmetry, there are also indirect ways how one can notice its influence.
The quadrupole contribution alters the field of the vortex electron at the times $t \lesssim t_d \gg t_c$.
Therefore, the spectrum of the field in the rest frame gets modified at the following frequencies:
\begin{eqnarray}
&& \displaystyle
\omega_0 \gtrsim \omega_d = m \left (\frac{\lambda_c}{\sigma_{\perp}(0)}\right )^2 \ll m,
\label{omega}
\end{eqnarray}
which are much lower than the electron's rest energy, even though in the laboratory frame we might have a $\gamma = \varepsilon/m$-enhancement, $\omega \sim \gamma \omega_0 > \omega_0$, due to the Doppler effect.
As a result, a wide variety of the quasi-classical emission processes with the vortex electrons can be influenced by this non-paraxial contribution at the relatively low frequencies. 
Such processes embrace the radiation in external electromagnetic fields (bremsstrahlung, synchrotron radiation, etc.) and in matter (Cherenkov radiation, transition radiation, Smith-Purcell radiation, etc.).
For instance, for an electron beam of the width $\langle \rho(0) \rangle \sim 1$ nm, we have
$$
\omega_d \sim 10^{-2}\, \text{eV},\ \lambda_d \sim 100\, \mu\text{m},
$$
whereas for $\langle \rho(0) \rangle \sim 10$ nm the corresponding frequency is two orders of magnitude lower and $\lambda_d \sim 1$ cm. 

Thus, for tightly focused twisted beams of $\langle \rho(0)\rangle \sim 0.1 - 10$ nm the non-paraxial contribution results in a noticeable modification of the radiation spectrum 
at the frequencies in the range of (in the rest frame)
\begin{eqnarray}
&& \displaystyle
\omega_0 \sim 10^{-4} - 1\, \text{eV}.
\label{omegarange}
\end{eqnarray}
One of the efficient ways for the generation of intense radiation in the THz and millimeter range is the so-called Smith-Purcell mechanism of radiation \cite{SP},
arising when an electron passes nearby a grating. The spreading of the OAM-less Gaussian packet does not change the radiation characteristics much, 
whereas the vortex electron packet loses the azimuthal symmetry in time because of the corresponding quadrupole moment. 
This is likely to result in an azimuthal asymmetry of the Smith-Purcell radiation, which would reveal itself at the frequencies (\ref{omegarange}).
For these frequencies to be present in the radiation spectrum, the grating should be longer than $\lambda_d \sim 1$ cm, which is easily realizable. 
In order to make quantitative estimates of these non-paraxial effects, detailed calculations are needed. 

\

I am grateful to K.~Bliokh, I.~Ivanov, P.~Kazinski, V.~Serbo, A.~Tishchenko, and A.~Zhevlakov for useful discussions and criticism. 
This work was supported by the Russian Science Foundation (Project No.\,17-72-20013).

\end{document}